\documentstyle[preprint,aps,epsfig,floats]{revtex}
\def\laq{\ \raise 0.4ex\hbox{$<$}\kern -0.8em\lower 0.62
ex\hbox{$\sim$}\ }
\def\gaq{\ \raise 0.4ex\hbox{$>$}\kern -0.7em\lower 0.62
ex\hbox{$\sim$}\ }

\def\PRD{{\em Phys. Rev.} D}

\begin{document}

 \preprint{\vbox{\baselineskip=12pt
 \rightline{IFUP-TH 30/99}
 \vskip1truecm}}

\title{Generalized Second Law in String Cosmology}

\author{Ram Brustein${}^{(1)}$, Stefano Foffa${}^{(2)}$,
and Riccardo Sturani${}^{(3),(2)}$}
\address{
 (1) Department of Physics,
 Ben-Gurion University, Beer-Sheva 84105, Israel\\
 (2) Dipartimento di Fisica, via Buonarroti 2, I-56100, Pisa, Italy\\
 and INFN, sezione di Pisa\\
 (3) Scuola Normale Superiore, piazza dei Cavalieri, I-56125, Pisa, Italy\\
 email: ramyb@bgumail.bgu.ac.il, foffa@ibmth.df.unipi.it, sturani@cibs.sns.it}

\maketitle

\begin{abstract}
A generalized second law  in string cosmology  accounts for
geometric and quantum entropy in addition to ordinary sources of
entropy. The proposed generalized second law forbids singular
string cosmologies, under certain conditions,
 and forces a  graceful exit transition from
dilaton-driven inflation by bounding curvature and dilaton kinetic
energy.
 \end{abstract}

String theory is  a consistent theory of quantum gravity, with the
power to describe high curvature regions of space-time
\cite{Polchinski}, and as such we could expect it to teach us
about the fate of cosmological singularities, with the expectation
that singularities are smoothed and turned into brief epochs of
high curvature. However, many attempts to seduce an answer out of
string theory regarding cosmological singularities have failed so
far, even after the wave of recent new developments and results
\cite{Polchinski2}.
The reason is probably that most recent technical advancements in
string theory rely heavily on supersymmetry, but generic time
dependent solutions break all supersymmetries, and therefore known
methods are less powerful when applied to cosmology. We propose to
turn to general thermodynamical considerations in the quest to
understand cosmological singularities, as first suggested by
Bekenstein \cite{Bekenstein:1989wf} in the context of Einstein's
general relativity.  We propose that entropy considerations, in
particular accounting for geometric and quantum entropy,
accompanied by a generalized second law (GSL) demanding that
entropy never decreases, be added to supplement string theory, and
show that under certain conditions GSL forbids cosmological
singularities. The proposed GSL is different from GSL for black
holes \cite{gslbh}, but the idea that in addition to normal
entropy other sources of entropy have to be included has some
similarities.

The idea that entropy and thermodynamics can play a role in
cosmology has been stressed recently by Fischler and Susskind, who
suggested that the holographic principle \cite{holo}  may be
useful in constraining cosmological solutions
\cite{Fischler:1998st}, followed by several other investigations
in which their suggestion was critically examined or improved
\cite{Bak:1998vj}. Our discussion is closer to that of
\cite{Easther:1999gk}, and in particular to that of
\cite{Veneziano:1999ts} where a new Hubble entropy bound (HEB) was
formulated, and its  relevance to the analysis of cosmological
singularities in string cosmology pointed out. The idea that
geometric and quantum entropy should be added, and be accompanied
by GSL was introduced in \cite{Brustein:1999ua}.

We will focus on two sources of entropy. The first source is
 geometric entropy $S_g$, whose origin
is the existence of a cosmological horizon
\cite{Gibbons:1977mu,Srednicki:1993im,Brustein:1999ua}. Geometric
entropy has been calculated for special systems, but we assume
that it is a general property of a system with a cosmological
horizon, resulting from the existence of causal boundaries in
space-time. The concept of geometric entropy is closely related to
the holographic principle, and it has appeared in this connection
recently in  discussion of cosmological entropy bounds. For a
system with a cosmological horizon, geometric entropy within a
Hubble volume
 is given roughly, ignoring numerical factors, by the
area of the horizon. The second source is quantum entropy $S_{q}$,
associated with quantum fluctuations. Changes in $S_q$ take into
account ``quantum leakage" of entropy, resulting from the well
known phenomenon of freezing and defreezing of quantum
fluctuations. For example, quantum modes whose wavelength is
stretched by an accelerated cosmic expansion to the point that it
is larger than the horizon, become frozen (``exit the horizon"),
and are lost as dynamical modes, and conversely quantum modes
whose wavelength  becomes smaller than the Hubble length during a
period of decelerated expansion, thaw (``reenter the horizon") and
become dynamical again. This form of entropy was discussed in
\cite{prigogine,Brandenberger:1992sr}. Obviously, there is a
relation, which we will expose more fully, between quantum and
geometric entropy, since the rate of expansion of the universe
determines the cosmological horizon and the rate of change of
quantum entropy.

We adopt the definition of the total entropy of a domain
containing more than one cosmological horizon
\cite{Veneziano:1999ts}, for a given scale factor $a(t)$, and a
Hubble parameter $H(t)=\dot a/ a$, the number of cosmological
horizons within a given comoving volume $V=a(t)^3$ is given by the
total volume divided by the volume of a single horizon,
$n_H={a(t)^3}/{ |H(t)|^{-3}}$.  If the entropy within a given
horizon is $S^H$, then the total entropy is given by $ S=n_H S^H$.
We will ignore numerical factors, use units in which $c=1$,
$\hbar=1$,  $G_{N}=e^{\phi}/16\pi$, $\phi$ being the dilaton, and
discuss only flat, homogeneous, and isotropic string cosmologies
in the so-called string frame, in which the lowest order effective
action is
 ${\cal S}_{lo}=\int d^4 x \sqrt{-g} e^{-\phi}
 \left[R + \left(\partial\phi\right)^2\right]$.

In ordinary cosmology, geometric entropy within a Hubble volume
is given by its area $S_g^H=H^{-2} G_N^{-1}$, and therefore
specific geometric entropy is given by
 $s_{g}=|H| G_{N}^{-1}$ \cite{Brustein:1999ua}. A possible expression for
specific geometric entropy in string cosmology is obtained by
substituting $G_{N}=e^{\phi}$, leading to
\begin{equation}\label{sst}
s_{g}=|H| e^{-\phi}.
\end{equation}
Reassurance that $s_g$ is indeed given by (\ref{sst}) is provided
by the following observation. The action ${\cal S}_{lo}$ can be
expressed in a $(3+1)$ covariant form, using the 3-metric
$g_{ij}$, the extrinsic curvature $K_{ij}$, considering only
vanishing $3-$Ricci scalar and
homogeneous dilaton,
$
{\cal S}_{lo}=\int d^3 x dt \sqrt{g_{ij}} e^{-\phi}
 \left[ - 3 K_{ij}K^{ij} - 2 g^{ij}\partial_t{K_{ij}}
 + K^2 - (\partial_t {\phi})^2 \right].
$
Now, ${\cal S}_{lo}$  is invariant
under the symmetry transformation
$
g_{ij}\rightarrow e^{2 \lambda}g_{ij}$,
 $\phi \rightarrow \phi +3\lambda$, for an arbitrary time
 dependent $\lambda$.
From the variation of the action $\delta {\cal S}=\int d^3 x dt
\sqrt{g_{ij}} e^{-\phi} 4 K \dot{\lambda}$, we may read off the
current and conserved charge $Q=4 a^3 e^{-\phi} K$. The  symmetry
is exact in the flat homogeneous case, and it seems plausible that
it is a good symmetry even when  $\alpha '$ corrections are
present \cite{Gasperini:1997fu}. With definition (\ref{sst}),
the total geometric entropy
$
S_{g}=a^3 |H| e^{-\phi},
$
is proportional to the corresponding conserved charge. Adiabatic
evolution, determined by $\partial_t S_g=0$, leads to a familiar
equation,
$
\frac{\dot{H}}{H}-\dot{\phi}+3H= 0,
$
satisfied by the $(\pm)$ vacuum branches of string cosmology.

Quantum entropy for a single field in string cosmology is, as in
\cite{Brandenberger:1992sr,Brustein:1999ua}, given by
\begin{equation} \label{squantum}
 s_{q}=\int_{k_{min}}^{k_{max}} d^3k f(k)\, ,
\end{equation}
where for large occupation numbers $f(k)\simeq \ln n_k$.
The ultraviolet
cutoff $k_{max}$ is assumed to remain constant at the string
scale. The infrared cutoff $k_{min}$ is determined by the
perturbation equation
$
 \psi_{k_{c}}''+\left(k_{c}^2- \frac{\sqrt{s(\eta)} ''}{\sqrt{s(\eta)}}\right)
\psi_{k_{c}}=0,
$
where $\eta$ is conformal time $'=\partial_\eta$,
and $k_{c}$ is the comoving momentum related to
physical momentum $k(\eta)$ as $k_{c}=a(\eta) k(\eta)$. Modes for which
$k_{c}^2 \laq \frac{\sqrt{s}''}{\sqrt{s}}$ are ``frozen", and are
lost as dynamical modes. The ``pump field"
$s(\eta)=a^{2m} e^{\ell\phi}$, depends on
the background evolution and
 on the spin and dilaton coupling of various fields \cite{Brustein:1998wt}.
 We are interested in solutions for which
$a'/a\sim \phi' \sim 1/\eta$, and therefore, for all particles
$\frac{\sqrt{s}''}{\sqrt{s}}\sim 1/\eta^2$. It follows that
$k_{min}\sim H$. In other phases of cosmological evolution our
assumption does not necessarily hold, but in standard radiation
domination (RD) with frozen dilaton all modes reenter the horizon.
Using the reasonable approximation $f(k)\sim$ constant, we obtain,
as in \cite{Brustein:1999ua},
\begin{equation} \label{dtsquan}
 \Delta S_{q}\simeq=-\mu \Delta n_H.
\end{equation}
Parameter $\mu$ is positive, and  in many cases  proportional to
the number of species of particles, taking into account all degrees of
freedom of the system, perturbative and non-perturbative. The main
contribution to $\mu$ comes from light degrees of freedom and therefore
if some non-perturbative objects, such as D branes become light they
will make a substantial contribution to $\mu$.

We now turn to the  generalized second law of thermodynamics,
taking into account geometric and quantum entropy. Enforcing
$dS\ge 0$, and in particular,
 $\partial_t S= \partial_t S_g + \partial_t S_q\ge 0$,
 leads to an important inequality,
\begin{equation} \label{eq:gsl}
 \left(H^{-2} e^{-\phi}-\mu\right)\partial_t n_H+
 n_H \partial_t \left(H^{-2} e^{-\phi}\right) \ge 0.
\end{equation}
When quantum entropy is negligible compared to geometric entropy,
GSL (\ref{eq:gsl}) leads to
\begin{equation}\label{dotphibound}
\dot{\phi}\leq \frac{\dot{H}}{H}+3H,
\end{equation}
yielding a bound on $\dot{\phi}$, and therefore on dilaton kinetic
energy,  for a given $H$, $\dot{H}$. Bound (\ref{dotphibound}) was
first obtained in \cite{Veneziano:1999ts}, and interpreted as
following from a saturated HEB.

When quantum entropy becomes relevant we obtain another bound. We
are interested in a situation in which the universe expands,
$H>0$, and $\phi$ and $H$ are non-decreasing, and therefore
 $\partial_t \left(H^{-2} e^{-\phi}\right)\le 0$ and
 $\partial_t n_H>0$. A necessary condition for GSL to hold is that
\begin{equation} \label{eq:hbound}
 H^{2} \le \frac{ e^{-\phi}}{\mu},
\end{equation}
bounding total geometric entropy $H e^{-\phi}\leq
\frac{e^{-\frac{3}{2} \phi}}{\sqrt{\mu}}$. A bound similar to
(\ref{eq:hbound}) was obtained in \cite{Maggiore:1998cz}
from a holographic bound on the rate of production of D0 branes,
and in \cite{Veneziano:1999ts} by considering
entropy of reentering quantum fluctuations. We stress that to be
useful in analysis of cosmological singularities (\ref{eq:hbound})
has to be considered for perturbations that exit the horizon. We
note that if the condition (\ref{eq:hbound}) is satisfied, then
the cosmological evolution never reaches the nonperturbative
region described in \cite{Maggiore:1998cz}, allowing a
self-consistent analysis using the low energy effective action
approach.

It is not apriori clear that the form of GSL and entropy sources
remains unchanged when curvature becomes large, in fact, we may
expect higher order corrections to appear. For example, the
conserved charge of the scaling symmetry of the action will depend
in general on higher order curvature corrections. Nevertheless, in
the following we will assume that  specific geometric entropy is
given by eq.~(\ref{sst}), without higher order corrections, and
try to verify that, for some reason yet to be understood, there
are no higher order corrections to eq.~(\ref{sst}). Our results
are consistent with this assumption.

We turn now to apply our general analysis to the `pre-big-bang'
(PBB) string cosmology scenario \cite{pbb}, in which the universe
starts from a state of very small curvature and string coupling
and then undergoes a long phase of dilaton-driven inflation (DDI),
joining smoothly at later times  standard
 RD cosmology,  giving rise to a singularity
free inflationary cosmology. The high curvature phase joining DDI
and RD phases is identified with the `big bang' of standard
cosmology. A key issue confronting this scenario is whether, and
under what conditions, can the graceful exit transition from DDI
to RD be completed \cite{Brustein:1994kw}. In particular, it was
argued that curvature is bounded by an algebraic fixed point
behaviour  when both $H$ and $\dot\phi$ are constants and the
universe is in a linear-dilaton deSitter space
\cite{Gasperini:1997fu}, and coupling is bounded by quantum
corrections \cite{Brustein:1997ny,Foffa:1999dv}. But it became
clear that another general theoretical ingredient is missing, and
we propose that GSL is that missing ingredient.

We have studied numerically  examples of PBB string cosmologies to
verify that the overall picture we suggest is valid in cases that
can be analyzed explicitly. We first consider, as in
\cite{Gasperini:1997fu,Brustein:1999yq}, $\alpha'$ corrections to
the lowest order string effective action,
\begin{equation}\label{eq2}
{\cal S}=\frac{1}{16 \pi \alpha '}\int d^4 x \sqrt{-g}
e^{-\phi}\left[R + \left(\partial\phi\right)^2 + \frac{1}{2}{\cal
L}_{\alpha '}\right]\, ,
\end{equation}
where
\begin{eqnarray}\label{alfa}
 {\cal L}_{\alpha '}=k \alpha' \Biggl[&& \frac{1}{2}R^2_{GB} + A
\left(\partial\phi\right)^4 +
D \partial^2\phi\left(\partial\phi\right)^2 \nonumber \\
&& +C\left(R^{\mu\nu}-\frac{1}{2}g^{\mu\nu}R\right)
\partial_{\mu}\phi\partial_{\nu}\phi\Biggr],
\end{eqnarray}
with $C=-(2 A + 2D +1)$, is the most general form of four
derivative corrections that lead to equations of motion with at
most  second (time) derivatives. The rationale for this choice was
explained in \cite{Brustein:1999yq}.  $k$ is a numerical factor
depending on the type of string theory. Action (\ref{eq2}) leads
to equations of motion,
$
- 3 H^2 + \dot{\bar{\phi}}^2- \bar{\rho}=0
$,
$
\bar{\sigma} - 2 \dot{H} + 2 H \dot{\bar{\phi}}=0
$,
$
\bar{\lambda} - 3 H^2 - \dot{\bar{\phi}}^2 + 2 \ddot{\bar{\phi}}=0
$, where $\bar{\rho}$, $\bar{\lambda}$, $\bar{\sigma}$ are
effective sources parameterizing the contribution of $\alpha '$
corrections \cite{Brustein:1999yq}. Parameters $A$ and $D$ should
have been determined  by string theory, however, at the moment, it
is not possible to calculate them in general. If $A$, $D$  were
determined we could just use the results and check whether their
generic cosmological solutions are non-singular, but since $A$,
$D$ are unavailable at the moment, we turn to GSL to restrict
them.

\begin{figure}
\centerline{\psfig{figure=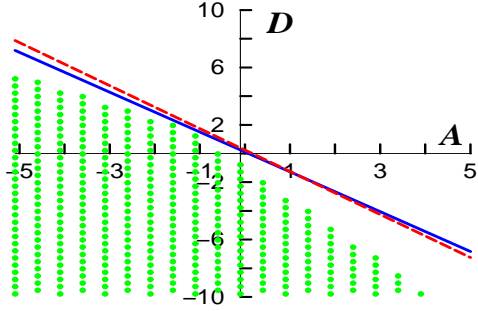,height=2.0in,width=3.2in}}
 \caption{ {\label{fig1}}
 {Two lines, separating actions whose generic solutions ``turn the right
way'' at the early stages of evolution (red-dashed),
 and actions whose generic solutions satisfy classical GSL
while close to the $(+)$ branch vacuum (blue-solid). The
dots represent  $(A,D)$ values whose generic solutions
reach a fixed point, and are all in the "allowed" region.} }
\end{figure}

First, we look at the initial stages of the evolution when the
string coupling and $H$ are very small. We find that not all the
values of the parameters $A$, $D$ are allowed by GSL. The
condition $\bar{\sigma}\geq0$, which  is equivalent to GSL on
generic solutions at the very early stage of the evolution, if the
only relevant form of entropy is geometric entropy, leads to the
following condition  on  $A$, $D$ (first obtained by R. Madden
\cite{dick}),
$
40.05 A + 28.86 D \leq 7.253.
$
The values of $A$, $D$ which satisfy this inequality are labeled
``allowed", and the rest are ``forbidden". In
\cite{Brustein:1999yq} a condition that $\alpha'$ corrections are
such that solutions start to turn towards a fixed point
at the very early stages of their evolution was found
 $ 61.1768 A + 40.8475 D \leq 16.083$,
and such solutions were labeled ``turning the right way". Both
conditions  are displayed in Fig.~\ref{fig1}. They select almost
the same region of $(A,D)$ space,  a gratifying result, GSL
``forbids" actions whose generic solutions are singular and do not
reach a fixed point.
\begin{figure}
\centerline{\psfig{figure=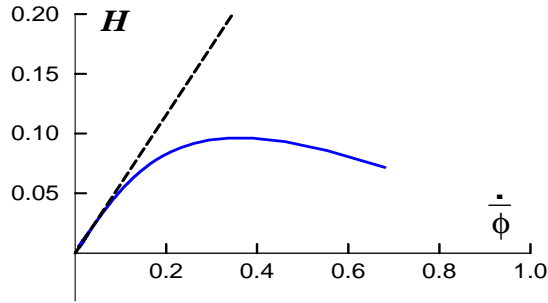,height=2.0in,width=3.2in}}
 \caption{ {\label{fig2}}
 {Typical solution that
 ``turns the wrong way". The dashed line is the $(+)$ branch vacuum.} }
\end{figure}
We further observe that
 generic solutions which ``turn the  wrong way"
at the early stages of their evolution continue their course in a
way similar to the solution presented in Fig.~\ref{fig2}. We find
numerically that at a certain moment in time $H$ starts to
decrease, at that point $\dot{H}=0$ and particle production
effects are still extremely weak, and therefore
(\ref{dotphibound}) is the relevant bound, but (\ref{dotphibound})
is certainly violated.

We have scanned the $(A,D)$ plane  to check whether a
generic solution that reaches a fixed point respects GSL
throughout the whole evolution, and conversely, whether a generic
solution obeying GSL evolves towards a fixed point. The results
are shown in Fig.~\ref{fig1}, clearly, the ``forbidden" region
does not contain actions whose generic solutions go to fixed
points. Nevertheless, there are some $(A,D)$  values located in
the small wedges near the bounding lines, for
which the corresponding solutions always satisfy
(\ref{dotphibound}), but do not reach a fixed point, and are
singular. This happens because they meet a cusp singularity.
Consistency requires adding higher order $\alpha'$ corrections when
cusp singularities are approached, which we will not attempt here.

If particle production effects are strong, the quantum part of GSL
adds  bound (\ref{eq:hbound}), which  adds another ``forbidden"
region in the $(H,\dot{\bar{\phi}})$ plane, the region above a
straight line parallel to the $\dot{\bar{\phi}}$ axis.
The quantum part of GSL has therefore a significant impact on corrections
to the effective action.   On a fixed point $\phi$ is
still increasing, and therefore the bounding line described by
(\ref{eq:hbound}) is moving downwards, and when the critical line
moves below the fixed point,  GSL is violated. This means that
when a certain critical value of the coupling $e^{\phi}$ is
reached, the solution can no longer stay on the fixed point, and
it must move away towards an exit. One way this can happen is
if quantum corrections, perhaps of the type discussed in
\cite{Brustein:1997ny,Foffa:1999dv} exist.

\begin{figure}
\centerline{\psfig{figure=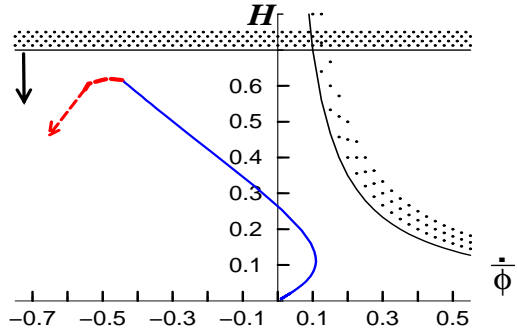,height=2.0in,width=3.2in}}
 \caption{ {\label{fig4}}
 {Graceful exit enforced by GSL on generic solutions.
 The horizontal line is bound (\ref{eq:hbound}) and the curve
on the right is bound (\ref{dotphibound}),  shaded
regions indicate  GSL violation. } }
\end{figure}
The full GSL therefore forces actions to have generic solutions
that are non-singular, classical GSL bounds
 dilaton kinetic energy and quantum  GSL bounds $H$
and therefore, at a certain moment of the evolution  $\dot{H}$
must vanish (at least asymptotically), and then curvature is
bounded. If  cusp singularities are removed by adding higher order
corrections, as might be expected,  we can apply GSL with similar conclusions
also in this case.
 A schematic graceful exit enforced by  GSL is shown in
Fig.~\ref{fig4}. We conclude that the use of thermodynamics and
entropy in string cosmology  provides  model independent tools to
analyze cosmological solutions which are not yet under full
theoretical control. Our result indicate that if we impose GSL, in
addition to equations of motion,  non-singular string cosmology is
quite generic.

\acknowledgements
 Work  supported in part by the  Israel Science
Foundation.  S. Foffa and R. Sturani wish to thank the department
of physics at BGU for hospitality. It is a pleasure to thank R.
Madden and G. Veneziano for useful discussions and helpful
suggestions and comments, and R. Easther for a useful discussion.


\begin{references}

\bibitem{Polchinski}
J.~Polchinski, {\em String Theory}, Cambridge University Press, 1998.

\bibitem{Polchinski2}
See, e.g., J.~Polchinski, hep-th/9812104.

\bibitem{Bekenstein:1989wf}
J.D.~Bekenstein,
Int. J. Theor. Phys. {\bf 28} (1989) 967.

\bibitem{gslbh}
J.D.~Bekenstein, \PRD 7, 2333 (1973) ; \PRD 9, 3292 (1974)


\bibitem{holo}
G.~'t Hooft, gr-qc/9310026;
L.~Susskind, J. Math. Phys. {\bf 36} (1995) 6377.


\bibitem{Fischler:1998st}
W.~Fischler and L.~Susskind,
hep-th/9806039.

\bibitem{Bak:1998vj}
D.~Bak and S.~Rey, hep-th/9811008;
 hep-th/9902173;
A.K.~Biswas, J.~Maharana and R.K.~Pradhan, hep-th/9811051;
R.~Dawid, Phys. Lett. {\bf B451} (1999) 19;
S.K.~Rama and T.~Sarkar, Phys. Lett. {\bf B450} (1999) 55;
 J.D.~Barrow, astro-ph/9903225;
 N.~Kaloper and A.~Linde, hep-th/9904120;
R.~Bousso, hep-th/9905177;
hep-th/9906022.

\bibitem{Easther:1999gk}
R.~Easther and D.A.~Lowe, Phys. Rev. Lett. {\bf 82} (1999) 4967.

\bibitem{Veneziano:1999ts}
G.~Veneziano, Phys. Lett. {\bf B454} (1999) 22; hep-th/9907012.

\bibitem{Brustein:1999ua}
R.~Brustein, gr-qc/9904061.


\bibitem{Gibbons:1977mu}
G.W.~Gibbons and S.W.~Hawking,
Phys. Rev. {\bf D15} (1977) 2738.

\bibitem{Srednicki:1993im}
M.~Srednicki, Phys. Rev. Lett. {\bf 71} (1993) 666.


\bibitem{prigogine} I. Prigogine {\em et al.},
{\em Gen. Rel. and Grav.}, 21, 767 (1989).

\bibitem{Brandenberger:1992sr}
R.~Brandenberger, V.~Mukhanov, T.~Prokopec, Phys. Rev. Lett. {\bf
69} (1992) 3606;
M.~Gasperini and M.~Giovannini, Phys. Lett. {\bf B301} (1993) 334.

\bibitem{Gasperini:1997fu}
M.~Gasperini, M.~Maggiore and G.~Veneziano, Nucl. Phys. {\bf B494}
(1997) 315.

\bibitem{Brustein:1998wt}
E.J.~Copeland, R.~Easther and D.~Wands,
Phys. Rev. {\bf D56} (1997) 874;
R.~Brustein and M.~Hadad, Phys. Rev. {\bf D57} (1998) 725;
A.~Buonanno, {\em et al.}, JHEP {\bf 01} (1998) 004.

\bibitem{Maggiore:1998cz}
M.~Maggiore and A.~Riotto, Nucl. Phys. {\bf B548} (1999) 427.


\bibitem{pbb}
G.~Veneziano, Phys. Lett. {\bf B265} (1991) 287;
M.~Gasperini and G.~Veneziano, Astropart. Phys. {\bf 1} (1993)
317;
A large collection of references on string cosmology can be found
at \texttt{http://www.to.infn.it/$\sim$ gasperin}.





\bibitem{Brustein:1994kw}
R.~Brustein and G.~Veneziano, Phys. Lett. {\bf B329} (1994) 429;
N.~Kaloper, R.~Madden and K.A.~Olive, Nucl. Phys. {\bf B452}
(1995) 677.

\bibitem{Brustein:1997ny}
R.~Brustein and R.~Madden, Phys. Lett. {\bf B410} (1997) 110;
 Phys. Rev. {\bf D57} (1998) 712.

\bibitem{Foffa:1999dv}
S.~Foffa, M.~Maggiore and R.~Sturani,
hep-th/9903008.

\bibitem{Brustein:1999yq}
R.~Brustein and R.~Madden,
JHEP {\bf 07} (1999) 006.
\bibitem{dick}
R. Madden, private communication.







\end{references}
\end{document}